\documentclass[twocolumn,aps,superscriptaddress,preprintnumbers,nofootinbib]{revtex4-1}

\usepackage[height=25cm,width=19cm,centering]{geometry}

\usepackage{amsmath,ascmac}
\usepackage{comment}
\usepackage{ifpdf}
\usepackage{xcolor}
\usepackage[utopia]{mathdesign}

\ifpdf
\usepackage{graphicx}
\usepackage[bookmarksopen,colorlinks=true,
linkcolor=dark-blue,citecolor=light-red,urlcolor=light-red]{hyperref}
\else
\usepackage[dvipdfmx]{graphicx}
\usepackage[dvipdfmx,bookmarksopen,colorlinks=true,
linkcolor=dark-blue,citecolor=light-red,urlcolor=light-red]{hyperref}
\fi

\definecolor{red}{rgb}{1,0,0}
\definecolor{light-red}{rgb}{0.941,0.196,0.196}
\definecolor{dark-blue}{rgb}{0,0,0.545}

\newcommand{\Mpl}{M_{\rm Pl}}
\newcommand{\abs}[1]{\left\vert {#1} \right\vert}
\newcommand{\cphi}{\varphi}
\newcommand{\der}{\partial}

\begin{document}


\title{
Electroweak Vacuum Metastability and Low-scale Inflation 
}

\author{Yohei Ema}
\affiliation{Department of Physics, Faculty of Science, The University of Tokyo}
\author{Kyohei Mukaida}
\affiliation{Kavli IPMU (WPI), UTIAS, University of Tokyo, Kashiwa, 277-8583, Japan}
\author{Kazunori Nakayama}
\affiliation{Department of Physics, Faculty of Science, The University of Tokyo}
\affiliation{Kavli IPMU (WPI), UTIAS, University of Tokyo, Kashiwa, 277-8583, Japan}

\begin{abstract}
\noindent
We study the stability of the electroweak vacuum in low-scale inflation models whose Hubble parameter
is much smaller than the instability scale of the Higgs potential.
In general, couplings between the inflaton and Higgs
are present, and hence we study effects of these couplings during and after inflation.
We derive constraints on the couplings between the inflaton and Higgs
by requiring that they do not lead to catastrophic electroweak vacuum decay,
in particular, via resonant production of the Higgs particles.
\end{abstract}

\date{\today}
\maketitle
\preprint{UT 17-21}
\preprint{IPMU 17-0090}


\section{Introduction}
\label{sec:intro}
\setcounter{equation}{0}

The Higgs potential may have
a deeper minimum than the electroweak (EW) vacuum 
once we assume that the Standard Model (SM) is valid up to a certain high-energy scale
given the current observational results of the SM parameters.
It does not mean any contradiction with the present universe since 
the allowed values of the SM parameters are likely to cause the metastable vacuum
where the lifetime of the EW vacuum far exceeds the age of the universe~\cite{Sher:1988mj,Arnold:1989cb,Anderson:1990aa,Arnold:1991cv,Espinosa:1995se,Isidori:2001bm,Ellis:2009tp,Bezrukov:2009db,EliasMiro:2011aa,Bezrukov:2012sa,Degrassi:2012ry,Masina:2012tz,Buttazzo:2013uya,Bednyakov:2015sca}.\footnote{
For the gravitational correction, 
see \textit{e.g.} Refs.~\cite{Isidori:2007vm,Branchina:2016bws,Rajantie:2016hkj,Salvio:2016mvj} 
and references therein.
}
Still, the existence of such a deeper minimum might cause problems in the early universe~\cite{Burda:2015isa,Grinstein:2015jda,Burda:2016mou,Tetradis:2016vqb,Gorbunov:2017fhq,Canko:2017ebb,Mukaida:2017bgd}, especially
during~\cite{Espinosa:2007qp,Lebedev:2012sy,Kobakhidze:2013tn,Fairbairn:2014zia,Enqvist:2014bua,Hook:2014uia,Herranen:2014cua,Kamada:2014ufa,Shkerin:2015exa,Kearney:2015vba,Espinosa:2015qea,
East:2016anr,Joti:2017fwe} 
and after inflation~\cite{Herranen:2015ima,Ema:2016kpf,Kohri:2016wof,Enqvist:2016mqj,Postma:2017hbk,Ema:2017loe}.
For instance, we can derive an upper bound on the inflation energy scale 
if there is no sizable coupling between the inflaton/the Ricci scalar and the Higgs during inflation.
Otherwise, the Higgs acquires superhorizon fluctuations which are large enough
to overcome the potential barrier during inflation.
Thus, we assume that the EW vacuum is indeed metastable,
and study its implications on dynamics during and after inflation in this paper.

Previous studies in this direction are performed mainly in the context of high-scale inflation.
The reason is that the Hubble parameter during inflation $H_\mathrm{inf}$ 
must be at least comparable to the instability 
scale of the Higgs potential $h_\mathrm{inst}$ ($\sim 10^{10}\,{\rm GeV}$ 
for the center values of the SM parameters)
for inflation to have nontrivial effects on the EW vacuum,
since otherwise superhorizon fluctuations 
during inflation are too small to overcome the potential barrier.
However, the situation completely changes 
once we consider dynamics after inflation.
After inflation, or during the inflaton oscillation epoch, 
the typical scale of the system is at least as large as 
the inflaton mass $m_\phi$.
Thus, as long as $m_\phi > h_\mathrm{inst}$,
even low-scale inflation that satisfies $h_\mathrm{inst} \gg H_\mathrm{inf}$
may threaten the metastable EW vacuum.
This is possible because low-scale inflation models typically yield $m_\phi \gg H_\mathrm{inf}$.

In this paper, we study dynamics of the Higgs during the inflaton oscillation epoch
for low-scale inflation models with $m_\phi > h_\mathrm{inst}$ and $m_\phi \gg H_{\rm inf}$.
In general, there are no reasons to suppress couplings between the inflaton and the Higgs.
If these couplings are sizable, 
a resonant production of the Higgs particles occurs due to the inflaton oscillation,
which is the so-called ``preheating'' phenomenon~\cite{Kofman:1994rk,Kofman:1997yn}.
The produced Higgs particles may force the EW vacuum to decay into the deeper minimum
through the negative Higgs self-coupling.
Thus we may obtain tight upper bounds on the couplings 
by requiring that the EW vacuum survives the preheating epoch.

Previous studies on the preheating dynamics of the EW vacuum
focused on 
high-scale inflation models~\cite{Ema:2016kpf,Kohri:2016wof,Enqvist:2016mqj,Postma:2017hbk,Ema:2017loe}
but there are some qualitative differences between high- and low-scale inflation models.
For low-scale inflation models, 
one significant complexity arises due to the tachyonic instability of the inflaton fluctuation itself
during the last stage of inflation and the subsequent inflaton oscillation epoch~\cite{Desroche:2005yt,Brax:2010ai,Antusch:2015nla,Antusch:2015vna}.
It can be efficient enough to break the homogeneity of the inflaton field before the Higgs field fluctuation develops.
Our purpose in this paper is to derive the upper bounds on the Higgs-inflaton couplings in low-scale inflation models
taking these effects into account.

This paper is organized as follows.
In Sec.~\ref{sec:setup}, we explain our setup. 
Since low-scale inflation models typically correspond to small field inflation models,
we concentrate on hilltop inflation models in this paper.
In Sec.~\ref{sec:inflation}, we briefly discuss
the dynamics of the Higgs during inflation for low-scale inflation models.
In Sec.~\ref{sec:preheating}, we study the preheating dynamics of the Higgs and inflaton itself,
and qualitatively discuss the feature of the whole system.
In Sec.~\ref{sec:sim}, we perform numerical simulations to derive bounds on the Higgs-inflaton couplings.
Finally, Sec.~\ref{sec:sum} is devoted to summary and discussions.

\section{Setup}
\label{sec:setup}

In this section, we summarize our setup.
We take the Lagrangian as
\begin{align}
	\mathcal{L} 
	&= 
	\frac{M_\text{Pl}^2}{2}R  - \frac{1}{2}\left(\partial \phi\right)^2 
	-\frac{1}{2}\left(\partial h\right)^2 
	- U(\phi, h),
	\label{eq:Lag}
\end{align}
where $\Mpl$ is the reduced Planck scale, $R$ is the Ricci scalar,
$\phi$ is the inflaton, and $h$ is the Higgs.\footnote{
	We consider only one degree of freedom for simplicity.
	The results change only logarithmically even if we consider the full SU(2) doublet.
} 
We assume that the inflaton is singlet under the SM gauge group,
and hence trilinear as well as quartic 
portal couplings between the inflaton and the Higgs are allowed in general.
Thus we take the following generic form for the potential:
\begin{align}
	U(\phi, h)
	&=
	V(\phi)
	+ \frac{\widetilde{\sigma}_{\phi h}}{2}\phi h^2
	+ \frac{\lambda_{\phi h}}{2}\phi^2 h^2
	+ \frac{m_h^2}{2} h^2 + \frac{\lambda_h}{4}h^4,
	\label{eq:pot}
\end{align}
where $V$ is the inflaton potential, 
$m_h^2$ is the bare mass of Higgs, 
and $\widetilde{\sigma}_{\phi h}$, $\lambda_{\phi h}$, and $\lambda_h$ are coupling constants.
Note that the inflaton can have some gauge charges other than SM, such as U(1)$_{\rm B-L}$.
In that case, $\phi$ should be regarded as a radial component of the complex scalar,
and $\widetilde \sigma_{\phi h} = 0$.
In this paper, however, we keep $\widetilde \sigma_{\phi h} \neq 0$
to make our discussion generic.
Also, although it is higher dimensional, the following term may be relevant:
\begin{align}
	\delta \mathcal L_\text{kin} = c_\text{kin}\frac{h^2}{M_\text{Pl}^2} \left( \der \phi \right)^2.
\end{align}
It can be sizable, for it respects the shift symmetry, $\phi \to \phi + \text{const}$.
We can also consider the non-minimal coupling between the Higgs and $R$.
We first omit these terms for simplicity, and discuss their effects at the end of this paper.

Below we explain each term in detail.

\subsection{Inflaton potential}

As a prototype of an inflaton potential for low-scale inflation, 
we consider the hilltop model~\cite{Linde:1981mu,Albrecht:1982wi,Barenboim:2013wra,Boubekeur:2005zm} (see Refs.~\cite{Kumekawa:1994gx,Izawa:1996dv,Asaka:1999jb,Senoguz:2004ky} for supergravity embeddings):
\begin{align}
	V(\phi) = \Lambda^4 \left[1 - \left(\frac{\phi}{v_\phi}\right)^n \right]^2,  \label{Vinf}
\end{align}
where $n > 2$ is an integer and $v_\phi > 0$ is the vacuum expectation value (VEV) of the inflaton 
at the minimum of its potential.
The inflaton mass around the minimum is
\begin{align}
	m_\phi = \frac{\sqrt 2 n \Lambda^2}{v_\phi}.
\end{align}
Since we are interested in small field inflation models,
we assume that $v_\phi \ll M_\text{Pl}$.
Otherwise, the model would be rather similar to high-scale inflation models.
Inflation takes place in the flat region of the potential: $\vert \phi \vert \ll v_\phi$.
Here and in what follows, we consider the field space of the positive branch: $\phi>0$.\footnote{
	A pre-inflation before the observed inflation can solve the initial condition problem of the hilltop inflation.
	If there exist a Hubble induced mass term during the pre-inflation
	and a small $\mathbb Z_2$ ($\phi \to - \phi$) breaking term,
	the initial condition is dynamically selected~\cite{Izawa:1997df}.
}
The Hubble parameter at the end of inflation $H_\mathrm{inf}$ 
is typically much smaller than $m_\phi$ in this case:
\begin{align}
	\frac{H_\mathrm{inf}}{m_\phi} \simeq \frac{v_\phi}{\sqrt{6} nM_\text{Pl}} \ll 1.
	\label{eq:Hubble_neg}
\end{align}

Using the standard technique to calculate the large-scale curvature perturbation~\cite{Liddle:2000cg},
one finds the scalar spectral index and tensor-to-scalar ratio as
\begin{align}
	n_s \simeq 1- \frac{2}{N}\frac{n-1}{n-2},~~~~r \simeq \frac{16n}{N(n-2)} \left[ \frac{1}{2N n (n-2)} \frac{v_\phi^2}{M_\text{Pl}^2} \right]^{\frac{n}{n-2}},
	\label{eq:ns_r}
\end{align}
where $N$ is the e-folding number of the cosmic microwave background (CMB) scale, 
which lies between $50$ and $60$ depending on the subsequent thermal history.
Thus the tensor-to-scalar ratio is negligibly small in small-field models with $v_\phi\ll M_\text{Pl}$.
The overall normalization of the curvature perturbation observed by 
the Planck satellite~\cite{Ade:2015lrj} implies
\begin{align}
	\mathcal P_\zeta \simeq 2.2\times 10^{-9}\simeq\frac{\left[2n((n-2)N)^{n-1}\right]^{\frac{2}{n-2}}}{12\pi^2}\frac{\Lambda^4}{(v_\phi^n M_\text{Pl}^{n-4})^{\frac{2}{n-2}}}.  \label{Planck}
\end{align}
It relates $\Lambda$ and $v_\phi$ and hence there is essentially one parameter left, which we take $v_\phi$ hereafter.

\begin{figure}[t]
\centering
 	\includegraphics[width=\linewidth]{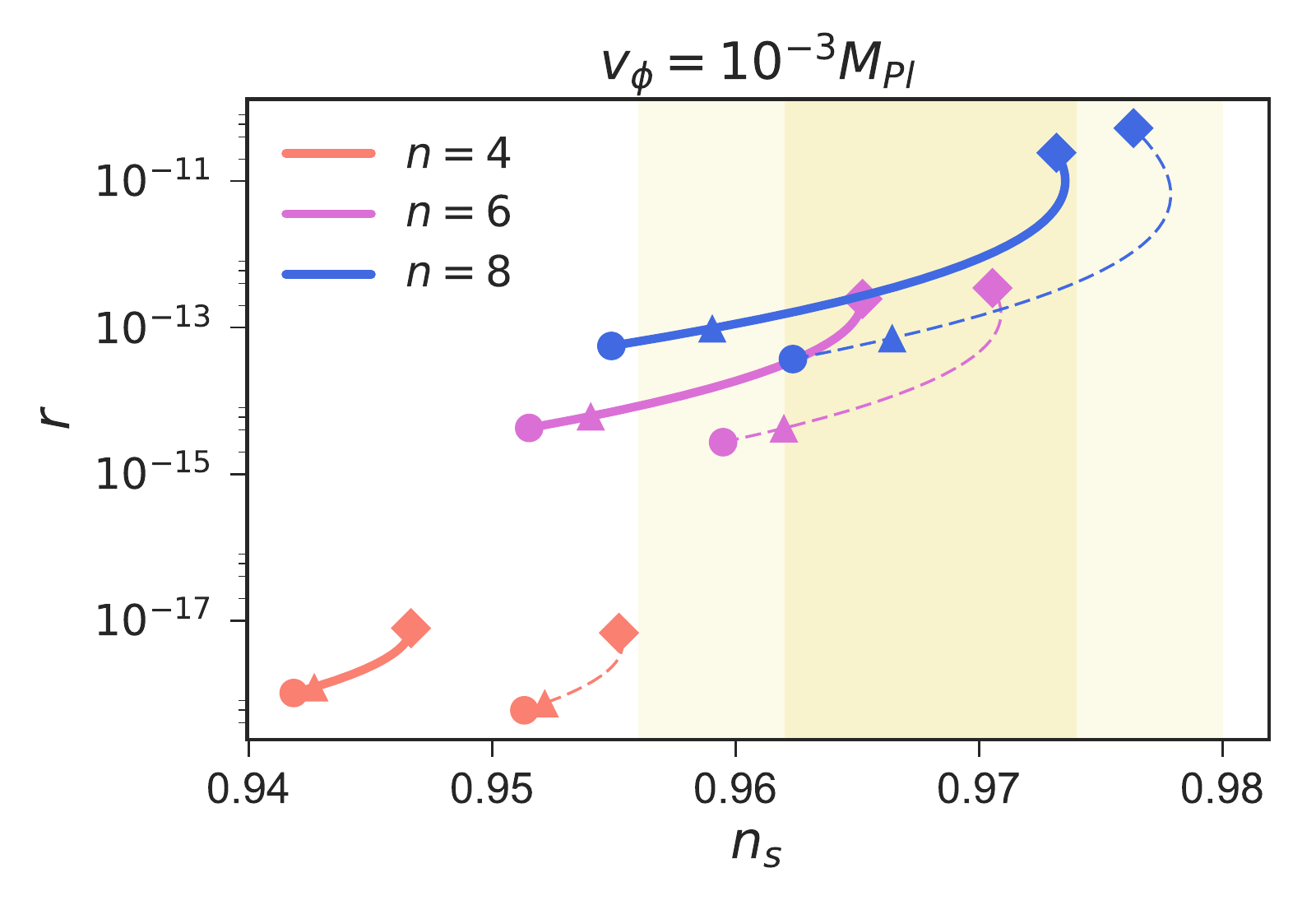}
	\caption{\small 
	Here we show $n_s$-$r$ plane for $n=4,6,8$ with varying $k$ from $10^{-4}$ to $10^{-2}$.
	The solid (dashed) lines correspond to $N=50$ $(60)$.
	The circle, triangle, and square represent points at $k = 10^{-4}, 10^{-3}, 10^{-2}$ respectively.
	The yellow shaded region stands for one and two sigma regions of $n_s$~\cite{Ade:2015lrj}.}
	\label{fig:ns}
\end{figure}

For a reasonable value of $n$, the predicted spectral index [Eq.~\eqref{eq:ns_r}] is slightly outside 
the favored range: $n_s = 0.968(6)$ at $68$\% confidence level~\cite{Ade:2015lrj}.
This discrepancy is resolved if there exists the following Planck suppressed operator~\cite{Kumekawa:1994gx,Izawa:1996dv}:
\begin{align}
	\delta V_\text{Pl} = - \Lambda^4 \frac{k}{2} \frac{\phi^2}{M_\text{Pl}^2},
\end{align}
with $k \lesssim \mathcal O (1/nN)$.
While it is too small to change the inflaton dynamics significantly,
it can shift the slow-roll parameter $\eta$ for a certain range of $k$.
If $n \geqslant 6$, it is possible to shift the spectral index within $68$\% confidence level
for $N = 50$--$60$.
See Fig.~\ref{fig:ns} and Ref.~\cite{Nakayama:2012dw}.
Since the suitable value of $k$ is small, this term is safely neglected in the oscillation phase.
Thus, we use the potential given in Eq.~\eqref{Vinf} in the following discussion.

\subsection{Higgs-inflaton couplings and bare mass term}
If we denote $\cphi \equiv v_\phi - \phi$,
the potential is given as
\begin{align}
	U(\phi, h) 
	=& 
	V(v_\phi - \cphi) 
	+ \frac{1}{2}\left(m_h^2 
	+ \widetilde{\sigma}_{\phi h}v_\phi 
	+ \lambda_{\phi h}v_\phi^2\right)h^2 \nonumber \\
	&+ \frac{\sigma_{\phi h}}{2} \cphi h^2 
	+ \frac{\lambda_{\phi h}}{2}\cphi^2 h^2
	+ \frac{\lambda_h}{4}h^4,
\end{align}
where we have defined 
\begin{align}
	\sigma_{\phi h} \equiv -\left(\widetilde{\sigma}_{\phi h} + 2 \lambda_{\phi h} v_\phi\right).
\end{align}
Note that $\cphi = 0$ at the minimum of the potential.
Here comes our crucial observation. 
In order to realize the EW scale, the bare Higgs mass 
and the mass coming from the inflaton VEV must be canceled:\footnote{
	We have neglected the EW scale since we are interested in the phenomena
	whose energy scale is much higher than the EW scale.
}
\begin{align}
	m_h^2 + \widetilde{\sigma}_{\phi h}v_\phi + \lambda_{\phi h}v_\phi^2 = 0.
	\label{eq:cancel}
\end{align}
It is a tuning, but we cannot avoid it
since we assume that the SM is valid up to some high-energy scale 
aside from the inflaton sector. Thus, the potential is now given by
\begin{align}
	U(\phi, h) 
	=& 
	V(v_\phi - \cphi)
	+ \frac{\sigma_{\phi h}}{2} \cphi h^2 
	+ \frac{\lambda_{\phi h}}{2}\cphi^2 h^2
	+ \frac{\lambda_h}{4}h^4.
	\label{eq:pot_around_vac}
\end{align}
In particular, the Higgs is almost massless at $\cphi = 0$.

Now we discuss quantum corrections to the potential.
The Higgs-inflaton couplings modify/induce runnings of the Higgs four point coupling/inflaton self-interactions.
Here let us focus on radiative corrections to the inflaton self-interactions;
for the Higgs four point coupling, see the next Sec.~\ref{sec:higgs_pot}.
As one can infer from Eq.~\eqref{Vinf}, the potential for the low-scale inflation has to be extremely flat,
and hence only a small change might spoil the successful inflation.
Suppose that the effective potential around the vacuum $\langle \phi \rangle \sim v_\phi$
is given by Eq.~\eqref{eq:pot_around_vac} at the end of inflation
for some renormalization scale $\mu$.
We will take $\mu$ as the typical scale of the preheating dynamics ($\mu \gtrsim m_\phi$).
See Sec.~\ref{sec:higgs_pot} for more details.
We put bounds on the couplings defined at this scale 
since we are interested the preheating dynamics.
Now the question is whether or not inflaton self-interactions 
are radiatively induced for $\phi \to 0$ and spoil the inflation.
At the one-loop level, the radiative correction is given by the Coleman-Weinberg effective potential,
\begin{align}
	V_{\rm CW}(\phi) &= \frac{m_h^4(\phi)}{64\pi^2}
	\ln \left(\frac{m_h^2(\phi)}{\mu^2}\right),
\end{align}
where we define $m_h^2(\phi) \equiv m_h^2+\widetilde\sigma_{\phi h}\phi + \lambda_{\phi h}\phi^2$,
and the couplings are evaluated at the scale $\mu$.
We have assumed $m_h^2 (\phi) > 0$ during inflation.
Otherwise, the Higgs potential might be destabilized during inflation
(see Sec.~\ref{sec:inflation}).
In order not to change the tree-level inflaton potential too much during inflation, we need 
$|\partial V_{\rm CW}/\partial\phi| \lesssim | \der V / \der \phi |$.
It roughly indicates
\begin{align}
	\abs{ \sigma_{\phi h} } \lesssim m_\phi \left(\frac{v_\phi}{M_\text{Pl}} \right)^{\frac{n-1}{n-2}},
	~~~~~~ \abs{\lambda_{\phi h}} \lesssim \frac{m_\phi}{M_\text{Pl}} \left(\frac{v_\phi}{M_\text{Pl}} \right)^{\frac{1}{n-2}},  \label{CW}
\end{align}
for $\widetilde \sigma_{\phi h} \neq 0$.
For $\widetilde \sigma_{\phi h} \simeq 0$,
we have instead
\begin{align}
	\abs{\sigma_{\phi h}}
	 \lesssim m_\phi \frac{v_\phi}{M_\text{Pl}},
	 ~~~~~
	 \abs{\lambda_{\phi h}} \lesssim \frac{m_\phi}{M_\text{Pl}}.
\end{align}

\subsection{Higgs potential}
\label{sec:higgs_pot}
Finally, we discuss the Higgs quartic self-coupling $\lambda_h$.
In order to understand the high-energy behavior of $\lambda_h$, 
we must carefully consider the scalar threshold correction~\cite{Lebedev:2012zw,EliasMiro:2012ay}.
Once we neglect the Higgs-inflaton quartic coupling,
the potential at around the minimum is written as
\begin{align}
	U \simeq \frac{m_\phi^2}{2}\left(\cphi + \frac{\sigma_{\phi h}h^2}{2m_\phi^2} \right)^2
	+ \frac{1}{4}\left(\lambda_h -  \frac{\sigma_{\phi h}^2}{2m_\phi^2}\right)h^4.
	\label{eq:pot_threshold}
\end{align}
Thus the Higgs potential below the energy scale of $m_\phi$ is
\begin{align}
	V_\mathrm{SM}(h) = \frac{\lambda_\mathrm{SM}}{4}h^4, ~~
	\lambda_\mathrm{SM} = \lambda_h - \frac{\sigma_{\phi h}^2}{2m_\phi^2}.
\end{align}
It is clear that the quartic coupling $\lambda_\mathrm{SM}$
in the low-energy effective theory is different from $\lambda_h$.

Up to the energy scale of $m_\phi$, the running of $\lambda_\mathrm{SM}$ is
just that of the SM, and hence it turns to be negative at around $10^{10}\,\mathrm{GeV}$
according to the current center values of the top/Higgs masses.
For simplicity, we approximate it as
\begin{align}
	\lambda_\mathrm{SM} = -0.01 \times \mathrm{sgn} \left(\mu - h_\mathrm{inst}\right)
	~~\mathrm{for}~~
	\mu < m_\phi,
\end{align}
where $\mu$ is the energy scale of the system and $h_\mathrm{inst}$ is the instability
scale of the Higgs potential which we take $h_\mathrm{inst} = 10^{10}\,\mathrm{GeV}$.
If $m_\phi < h_\mathrm{inst}$, 
$\lambda_h$ is positive at least up to at around $\mu = h_\mathrm{inst}$.\footnote{
The potential can be even absolutely stable depending on $\sigma_{\phi h}^2/m_\phi^2$ and 
the sign of $\lambda_{\phi h}$~\cite{Lebedev:2012zw,EliasMiro:2012ay}.
} 
Thus, to overcome the potential barrier,
the Higgs dispersion must be enhanced as large as $\langle h^2 \rangle \gtrsim h_\text{inst}^2 > m_\phi^2$.
However, such an enhancement requires a large coupling with inflaton which
is likely to spoil the flatness of the inflaton potential (see Eq.~\eqref{CW}).\footnote{
	In fact, the hilltop model $(n=6)$ with $m_\phi < h_\text{inst} \sim 10^{10}$\,GeV 
	cannot have large resonance parameters
	because of Eq.~\eqref{eq:CW_pq}.
} 
Therefore in this paper, we concentrate on the opposite case: 
\begin{align}
	m_\phi > h_\mathrm{inst}.
	\label{eq:m_gtr_inst}
\end{align}
Then, by matching at $\mu = m_\phi$, the boundary condition for $\lambda_h$ is roughly given as
\begin{align}
	\left.\lambda_h\right\rvert_{\mu = m_\phi} 
	= 
	-0.01 + \left.\frac{\sigma_{\phi h}^2}{2m_\phi^2}\right\rvert_{\mu = m_\phi}.
\end{align}
If $\sigma_{\phi h}^2/m_\phi^2 \gtrsim 0.01$, 
it may significantly affect $\lambda_h$ so that 
it helps to stabilize the Higgs potential at the high-energy region.\footnote{
If $\lambda_{\phi h}$ is negative, the potential may not be absolutely stable anyway,
depending on the precise form of $V(\phi)$.
}
Thus, there may be another minimum at around $h \simeq m_\phi$ and $\varphi \simeq -\sigma_{\phi h}$
because of Eqs.~\eqref{eq:pot_threshold} and \eqref{eq:m_gtr_inst}, and it may affect the dynamics of the Higgs in the early universe.

Instead of being involved in such a complexity, 
in this paper we simply concentrate on the case
\begin{align}
	\frac{\sigma_{\phi h}^2}{m_\phi^2} \ll 0.01.
\end{align}
Then, we may approximate the quartic coupling as
\begin{align}
	\lambda_h = -0.01 \times \mathrm{sgn} \left(\mu - h_\mathrm{inst}\right).
	\label{eq:lam_h}
\end{align}
We take the renormalization scale as 
$\mu = \mathrm{max}\left(H_\mathrm{inf}, h\right)$ during inflation~\cite{Herranen:2014cua}, 
and $\mu = \mathrm{max}\left(H, \sqrt{\langle h^2\rangle}\right)$
during preheating. Here $H$ is the Hubble parameter, and 
$\langle h^2\rangle$ is the dispersion of the Higgs field.
Actually, as soon as the resonant Higgs production occurs,
the dispersion becomes $\langle h^2 \rangle \gtrsim m_\phi^2$,
and hence it dominates over the Hubble parameter.

\section{Higgs dynamics during inflation}
\label{sec:inflation}

Before studying the preheating stage,
we summarize the Higgs dynamics during inflation in this section.
As studied extensively~\cite{Espinosa:2007qp,Lebedev:2012sy,Kobakhidze:2013tn,Fairbairn:2014zia,Enqvist:2014bua,Hook:2014uia,Herranen:2014cua,Kamada:2014ufa,Shkerin:2015exa,Kearney:2015vba,Espinosa:2015qea,
East:2016anr,Joti:2017fwe}, the de-Sitter fluctuation of the Higgs field
may lead to the collapse of the vacuum during inflation if the inflation scale is too high.
It is instructive to see what happens if the inflation scale is so low that $H_\text{inf} \ll h_\text{inst}$.

In the present model, since $\phi \ll v_\phi$ during inflation, 
the Higgs potential during inflation is approximately given by
\begin{align}
	V(h)
	&\simeq
	\frac{m_h^2}{2}h^2 + \frac{\lambda_h}{4}h^4,
\end{align}
where the bare Higgs mass $m_h^2$ satisfies Eq.~\eqref{eq:cancel}.
There are two possibilities: $m_h^2 <0$ and $m_h^2 >0$.

First, let us consider the case of tachyonic Higgs during inflation: $m_h^2 < 0$, 
or $\widetilde{\sigma}_{\phi h}v_\phi + \lambda_{\phi h}v_\phi^2 > 0$.
In this case, the parameters must satisfy
\begin{align}
	\left\lvert \lambda_h \right\rvert h_\mathrm{inst}^2 
	&> 
	|m_h^2|,
	\label{eq:tac_inf}
\end{align}
since otherwise the potential decreases monotonically toward large $h$ and
the Higgs may roll down to the deeper minimum during inflation.
As long as Eq.~\eqref{eq:tac_inf} is satisfied, the EW vacuum is stable during inflation 
if the Hubble scale during inflation is low enough, \textit{i.e.}, $H_\mathrm{inf} \ll h_\mathrm{inst}$.
Otherwise, the de-Sitter fluctuation of the Higgs field is too large to stay at the local minimum of the potential.

Next, let us consider the opposite case: $m_h^2 > 0$, 
or $\widetilde{\sigma}_{\phi h}v_\phi + \lambda_{\phi h}v_\phi^2 < 0$.
In this case, $h= 0$ is always a local minimum of the potential, and it is stable against the de-Sitter fluctuation if
\begin{align}
	H^2_{\rm inf} \ll {\rm max}\left[ h^2_{\rm inst},~m_h^2/|\lambda_h| \right].
	\label{eq:stb_inf}
\end{align}

If the condition~\eqref{eq:tac_inf} or~\eqref{eq:stb_inf} is satisfied, 
the Higgs field effectively stays at around the origin without overshooting the potential barrier
due to the de-Sitter fluctuation.
However, it does not guarantee the vacuum stability \textit{after} inflation,
since the Higgs fluctuation can be resonantly enhanced during the preheating stage as studied in detail in the next section.

\section{Inflaton and Higgs dynamics during preheating}
\label{sec:preheating}

In this section, we analytically describe the preheating dynamics of our system.
We first discuss resonant inflaton production in Sec.~\ref{sec:inf}. 
Since the inflaton potential at around the minimum is far from 
quadratic in the low-scale inflation model, inflaton particles are resonantly produced from the inflaton condensation.
In fact, the inflaton particles can be even tachyonic during the preheating epoch. Hence, the inflaton production is so efficient that
the backreaction destroys the inflaton condensation within several times of the oscillation.
It sets the end of the preheating epoch, 
and hence sets the upper bound of the time we follow in this paper.

Then we discuss resonant Higgs production in Sec.~\ref{sec:Higgs}.
There we make use of a crude approximation that 
the inflaton potential is dominated by the quadratic one.
This is because the purpose of this subsection 
is to understand the Higgs production qualitatively
and to make an order of magnitude estimation of the constraints on the couplings.
More rigorous analysis is performed numerically in the next section.

\subsection{Inflaton dynamics during tachyonic oscillation}
\label{sec:inf}

The inflaton oscillation is typically dominated by the flat part of the potential just after inflation,
and it causes a so-called tachyonic preheating phenomenon.
Below we closely follow the discussion in Ref.~\cite{Brax:2010ai} concerning
the linear regime of the tachyonic preheating. More details are given in App.~\ref{sec:app}.

There are two stages of tachyonic preheating.
The first stage is further divided into the epoch between the point $|\eta| = 1$ and $\epsilon=1$,
and the interval between $\epsilon=1$ and the first passage of $\phi=v_\phi$.
Here $\epsilon$ and $\eta$ are the slow-roll parameters: 
$\epsilon\equiv M_\text{Pl}^2(V'/V)^2/2$, $\eta\equiv M_\text{Pl}^2 V''/V$.
The tachyonic growth starts after $|\eta| \gtrsim 1$, 
where there is a large hierarchy between $\eta$ and $\epsilon$ in low-scale inflation models.
Therefore, the tachyonic growth occurs at the plateau regime of the inflaton potential, and
the inflaton fluctuation with $k/a \lesssim H_{\rm inf}$ will develop.
While the inflaton is rolling down the potential, 
higher momentum modes with $H_{\rm inf} < k/a \lesssim m_\phi$ also experience tachyonic growth,
but modes with low $k/a$ $(\lesssim H_{\rm inf})$ are most enhanced because they have more time to develop.
The inflaton fluctuation with such low-momenta at $\phi=v_\phi$ is estimated as\footnote{
	More precisely, the inflaton fluctuation $\delta\phi_k$ should be regarded as its gauge-invariant generalization 
	taking account of the scalar metric perturbation (see Ref.~\cite{Brax:2010ai} for more detail).
	Also, note that the curvature perturbation on large-scale is conserved since $\delta\phi_k \propto \dot\phi$.
}
\begin{align}
	\frac{\delta\phi_k(\phi(t)=v_\phi)}{\delta\phi_k(|\eta|=1)}= \frac{\dot\phi(\phi(t)=v_\phi)}{\dot\phi(|\eta|=1)} 
	\sim \left( \frac{M_\text{Pl}}{v_\phi} \right)^{\frac{n}{n-2}}.
\end{align}
Then the condition for the inflaton fluctuation to remain perturbative after the first passage of $\phi(t)=v_\phi$ is
$\left<\delta\phi^2\right> \lesssim v_\phi^2$, and it leads to
\begin{align}
	\left( \frac{v_\phi}{M_\text{Pl}} \right)^{1/(n-2)} \gtrsim \frac{\Lambda}{v_\phi}.
\end{align}
Using the Planck normalization (\ref{Planck}), this translates into $v_\phi / M_\text{Pl} \gtrsim 10^{-6}$--$10^{-5}$ independently of $n$.
Otherwise, even within one inflaton oscillation, the inflaton condensate may be broken,
and the subsequent inflaton-Higgs dynamics would be too complicated.
To avoid this complexity, we focus on the case of $v_\phi / M_\text{Pl} \gtrsim 10^{-6}$--$10^{-5}$ 
so that we can reliably discuss the Higgs dynamics in the second stage explained below.

In the second stage, the system goes into tachyonic inflaton oscillation regime.
During this stage, the inflaton oscillation is far from harmonic 
because the most oscillation period is consumed at the flat part of the potential
$\phi \ll v_\phi$.
After the $j$-th oscillation of the inflaton, the field value at the lower endpoint is given by
\begin{align}
	\frac{\phi_j}{v_\phi} \simeq \left(\frac{j\sqrt 3}{2}\frac{v_\phi}{M_\text{Pl}}\right)^{1/n}.
\end{align}
The most enhanced mode during this tachyonic oscillation stage is basically determined by the curvature of the inflaton potential
at $\phi=\phi_j$:
\begin{align}
	\frac{k_{*}}{a} \simeq m_\phi\left(\frac{jv_\phi}{M_\text{Pl}}\right)^{(n-2)/(2n)}.
	\label{eq:p_inf}
\end{align}
It is this mode ($k=k_{*}$) that is most enhanced through the whole tachyonic preheating process.
Note that it is much different from the ordinary broad resonance in which the inflaton oscillates about the quadratic potential.
In our case, the fluctuation becomes nonlinear, \textit{i.e.}, $\langle \delta\phi^2\rangle \sim v_\phi^2$,
within several times of oscillation.
See App.~\ref{sec:app} for more detail.

In summary, the inflaton fluctuation becomes nonlinear within several times of oscillation 
due to the tachyonic preheating.
To avoid complications arising from the nonlinearity and thermalization as well as 
possible model dependent discussions, we conservatively require that 
the vacuum remains stable at least until the inflaton fluctuation becomes nonlinear in this paper. 
Otherwise, we cannot avoid the catastrophe anyway.
Thus, the tachyonic production of the inflaton particles sets 
the upper bound of the time during 
which we follow the dynamics in this paper.

\subsection{Higgs dynamics during preheating}
\label{sec:Higgs}

Now we are in a position to study the growth of the Higgs field fluctuation during the preheating stage.
In this subsection, we crudely approximate the inflaton potential as quadratic, 
although the actual inflaton potential just after inflation is typically far from quadratic
for low-scale inflation models.
Nevertheless, it helps us to understand the numerical results in the next section.

The potential of the inflaton and Higgs at the inflaton oscillation phase is 
\begin{align}
	U(\phi, h) 
	&= 
	\frac{m_\phi^2}{2}\cphi^2
	+ \frac{\lambda_h}{4}h^4
	+ \frac{\sigma_{\phi h}}{2} \cphi h^2 + \frac{\lambda_{\phi h}}{2}\cphi^2 h^2.
	\label{eq:pot_min}
\end{align}
The inflaton potential is approximately taken to be quadratic around the potential minimum.
We consider the preheating dynamics of this system, 
\textit{i.e.}, the resonant Higgs particle production
due to the inflaton oscillation.\footnote{
	The EW vacuum stability of this system during the preheating epoch  
	is studied for large field inflation models in Refs.~\cite{Enqvist:2016mqj,Ema:2017loe}.
}
The linearized equation of motion of the Higgs is
\begin{align}
	\ddot{h}_k + \left(k^2 + \sigma_{\phi h}\cphi + \lambda_{\phi h}\cphi^2\right)h_k = 0,
	\label{eq:eom_lin}
\end{align}
where the dot denotes the derivative with respect to the time.
We have moved to the momentum space with $k$ being the momentum,
and neglected the Hubble expansion 
because of Eq.~\eqref{eq:Hubble_neg}.
The inflaton oscillation is described as
\begin{align}
	\cphi = \cphi_\mathrm{ini} \cos\left(m_\phi t\right),
	\label{eq:inf_osc}
\end{align}
under the quadratic approximation. 
Here $\cphi_\mathrm{ini}$ is the initial inflaton oscillation amplitude, 
which is roughly $\cphi_\mathrm{ini} \sim v_\phi$
(remember that $\cphi \equiv v_\phi - \phi$).
Note again that,
although the oscillation amplitude is a time-decreasing function due to the Hubble expansion, 
the Hubble parameter is so small that the effect of Hubble expansion is practically negligible
in low-scale inflation models with $H_{\rm inf}\ll m_\phi$ [Eq.~\eqref{eq:Hubble_neg}].

\begin{figure}[t]
\centering
 	\includegraphics[width=.8\linewidth]{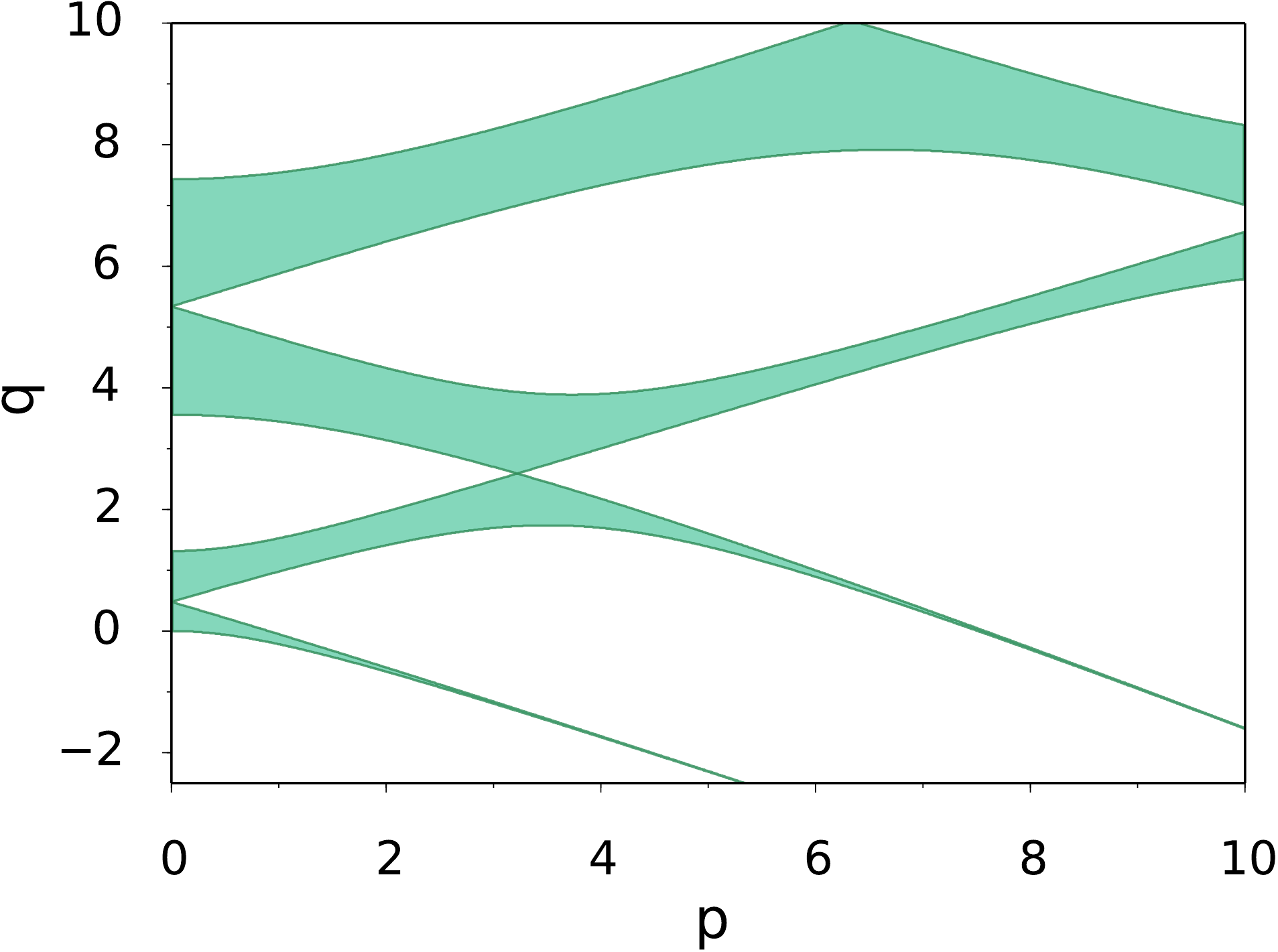}
	\caption{\small 
	The stability/instability chart of the Whittaker-Hill equation for the mode with $k = 0$.
	The green shaded region corresponds to the stability region, 
	while the unshaded region does to the instability region.
	The chart is even with respect to $p$.}
	\label{fig:WH}
\end{figure}

By substituting it to Eq.~\eqref{eq:eom_lin}, we obtain the Whittaker-Hill equation:
\begin{align}
	h_k'' + \left[A_k + 2p\cos2z + 2q \cos4z \right]h_k = 0,
	\label{eq:WH}
\end{align}
where 
\begin{align}
	A_k \equiv \frac{4k^2}{m_\phi^2} + 2q,~~
	p \equiv \frac{2\sigma_{\phi h}\cphi_\mathrm{ini}}{m_\phi^2},~~
	q \equiv \frac{\lambda_{\phi h}\cphi_\mathrm{ini}^2}{m_\phi^2},~~
	z \equiv \frac{m_\phi t}{2},
	\label{eq:res_param}
\end{align}
and the prime denotes the derivative with respect to $z$.
The term with $q$ leads to the usual parametric resonance~\cite{Kofman:1994rk,Kofman:1997yn}, 
while the term with $p$ potentially leads to the tachyonic resonance~\cite{Dufaux:2006ee}.
In Fig.~\ref{fig:WH}, we show the stability/instability chart of the Whittaker-Hill equation
for $k = 0$ for both positive and negative $q$.
If the parameters are in the instability region (the unshaded region), 
Eq.~\eqref{eq:WH} has exponentially growing solutions, resulting in the resonant Higgs production.
A similar stability/instability chart can be drawn for finite $k$ modes.
The resonance parameters $p$ and $q$ are useful for estimating the strength of the resonance
even for a potential that is far from quadratic, as in the case of the hilltop potential.
For more details on the Whittaker-Hill equation and the Floquet theory, see, \textit{e.g.}, 
Refs.~\cite{Lachapelle:2008sy,Enqvist:2016mqj} and references therein.

In terms of the resonance parameters, the condition
\begin{align}
	p + 2q \geq 0,
	\label{eq:Higgs_stb_pq}
\end{align}
is necessary for the Higgs not to be tachyonic during inflation.
Although it does not necessarily cause a problem 
even if the Higgs is tachyonic during inflation as long as Eq.~\eqref{eq:tac_inf} is fulfilled (see Sec.~\ref{sec:inflation}),
we will assume that Eq.~\eqref{eq:Higgs_stb_pq} holds in the following for simplicity.

Once the resonant Higgs production occurs,
it forces the EW vacuum to decay into 
the deeper minimum~\cite{Herranen:2015ima,Ema:2017loe,
Ema:2016kpf,Kohri:2016wof,Enqvist:2016mqj}.
This is because the produced Higgs particles induce 
the following tachyonic mass from the Higgs self-quartic coupling:
\begin{align}
	m_{\mathrm{tac};h}^2 \simeq 3\lambda_h \langle h^2\rangle,
\end{align}
where we have used the mean-field approximation. 
Note that the dispersion is typically $\langle h^2\rangle \gtrsim m_\phi^2$
for the resonant particle production, and thus we expect $\lambda_h < 0$ 
as can be seen from Eqs.~\eqref{eq:m_gtr_inst} and \eqref{eq:lam_h}.

Thus we can constraint the resonance parameters, or the couplings, by requiring that the EW vacuum is stable during the preheating
(or within several times of the inflaton oscillation).
The tachyonic resonance is effective if $\vert p\vert$ exceeds of order unity (see Fig.~\ref{fig:WH}), 
so we may require
\begin{align}
	\left\vert p \right\vert \lesssim \mathcal{O}(1),
	\label{eq:const_p_a}
\end{align}
for the EW vacuum stability during the preheating.
We will confirm this expectation 
by classical lattice simulations~\cite{Polarski:1995jg,Khlebnikov:1996mc} 
with a full hilltop inflaton potential in the next section.
Note that Eq.~\eqref{eq:const_p_a} implies that $\vert q \vert \lesssim \mathcal{O}(1)$
without any accidental cancellation between $\sigma_{\phi h}$ and $\lambda_{\phi h}$.
However, we will also discuss the case $\vert p\vert \lesssim \mathcal{O}(1)$ and 
$\vert q \vert \gg \mathcal{O}(1)$ at the end of the next section
for the completeness of this paper.\footnote{
	Note that $q \gtrsim -\mathcal{O}(1)$ from 
	Eqs.~\eqref{eq:Higgs_stb_pq} and~\eqref{eq:const_p_a}.
	\label{fn:q}
}

\section{Numerical simulation}
\label{sec:sim}

\begin{figure*}[t]
\centering
 	  \includegraphics[width=0.4\textwidth]{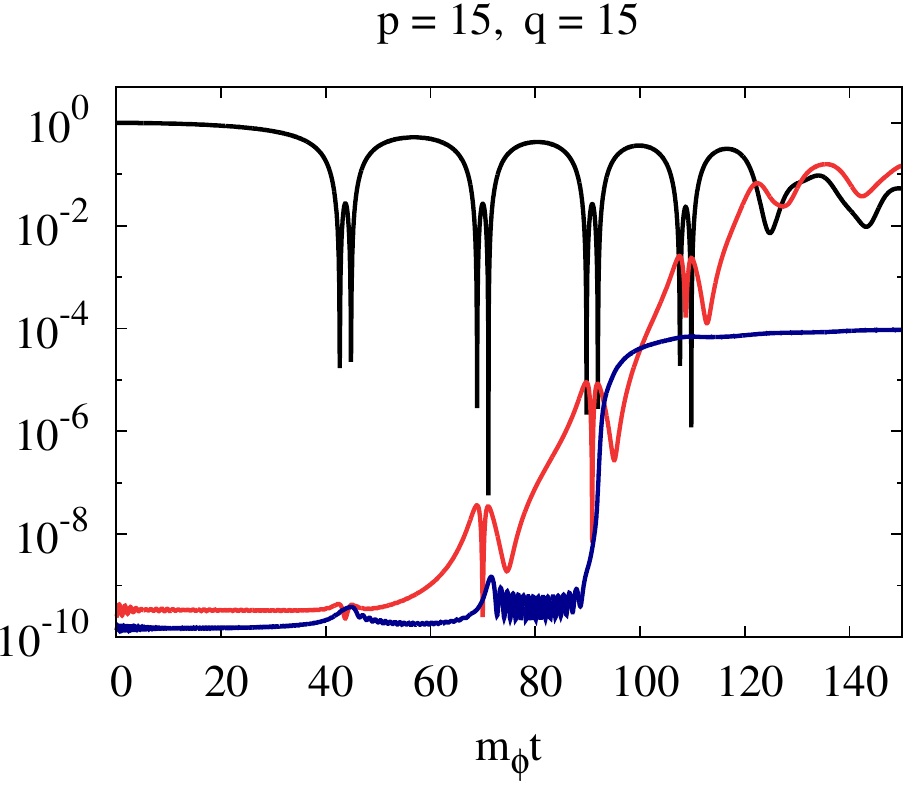} 
	  \hspace{15mm}
	  \includegraphics[width=0.4\textwidth]{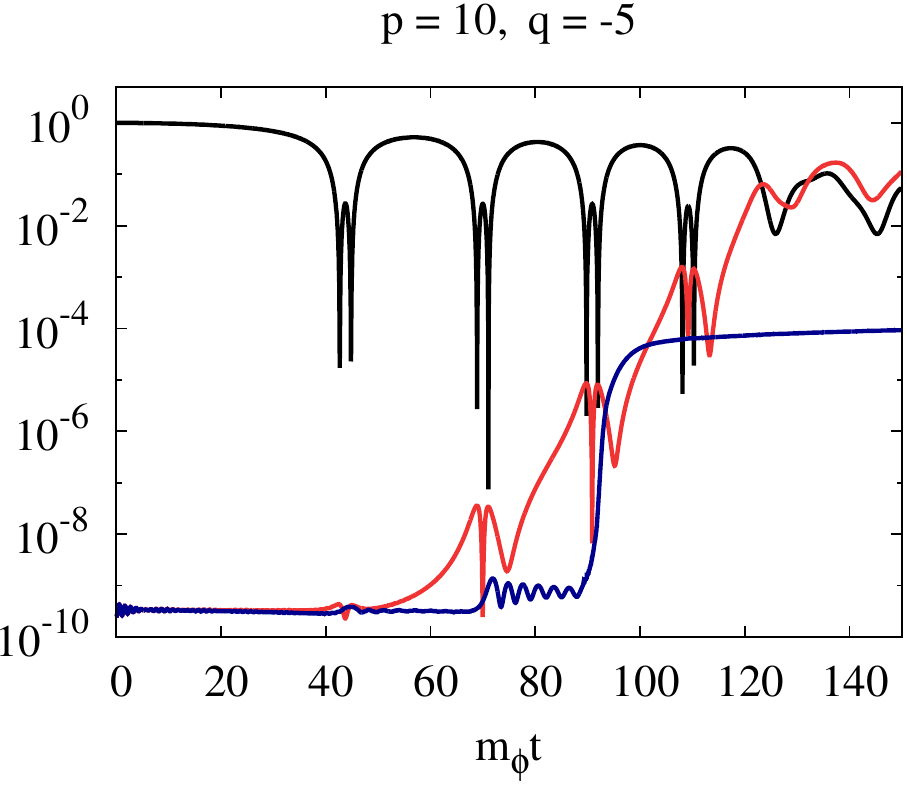}
	  \vspace{5mm}

	  \includegraphics[width=0.4\textwidth]{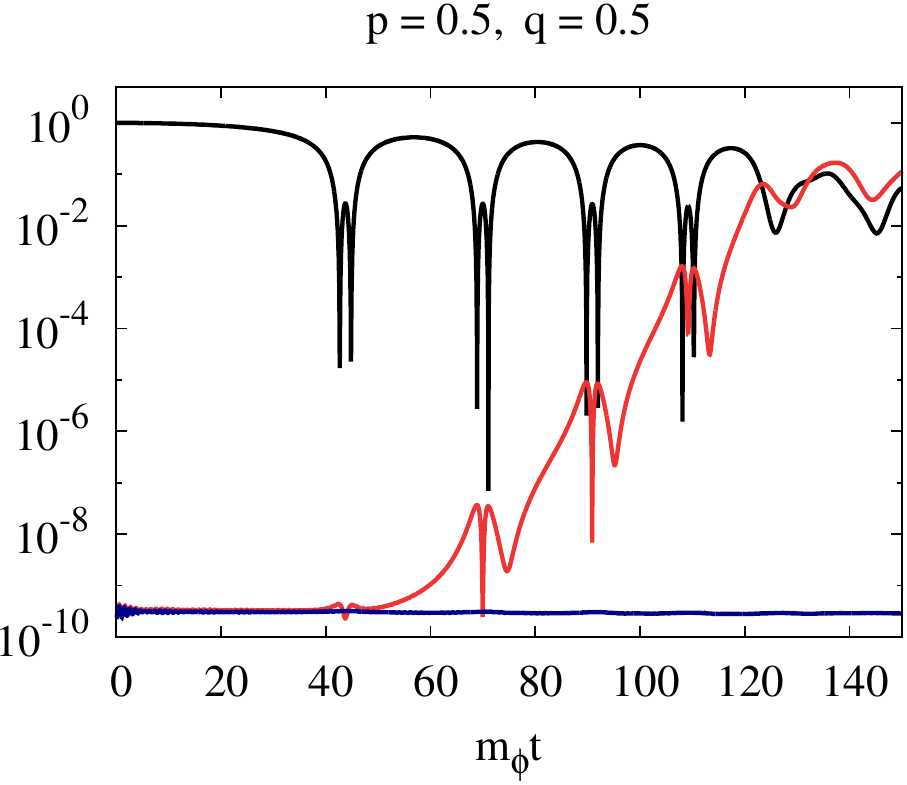}
	  \hspace{15mm}
	  \includegraphics[width=0.4\textwidth]{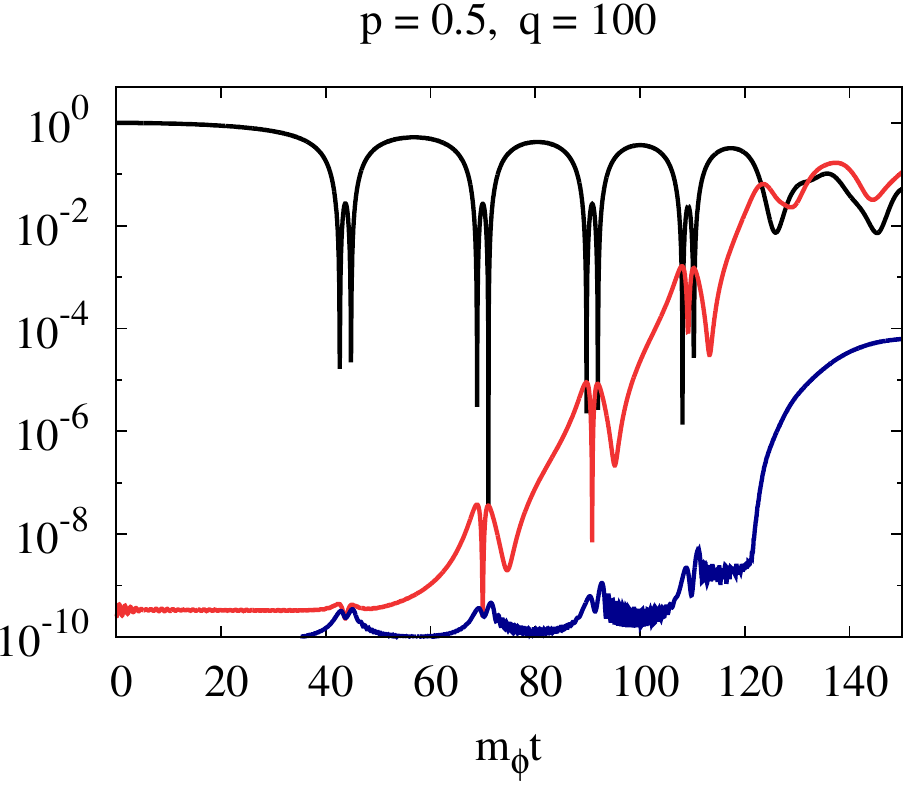}
\caption{\small
	The time evolution of the inflaton and the Higgs for $v_\phi = 10^{-2}\Mpl$ up to $m_\phi t = 150$. 
	The black line is the inflaton condensation $\langle \varphi \rangle^2$,
	\textcolor{light-red}{the red line} is the inflaton two point function 
	$\langle \varphi^2 \rangle - \langle \varphi \rangle^2$ and
	\textcolor{dark-blue}{the blue line} is the Higgs two point function $\langle h^2 \rangle$,
	where the angle brackets denote the spatial average.
	They are normalized by the initial inflaton amplitude $\varphi_\text{ini}$.
	The EW vacuum is stable for $(p, q) = (0.5, 0.5)$, while
	it is destabilized during the preheating for the other cases.
	The lower right panel corresponds to the case with an accidental cancellation
	between $\sigma_{\phi h}$ and $\lambda_{\phi h}$.
}
\label{fig:v10-2}
\end{figure*}
\begin{figure*}
\centering
 	  \includegraphics[width=0.4\textwidth]{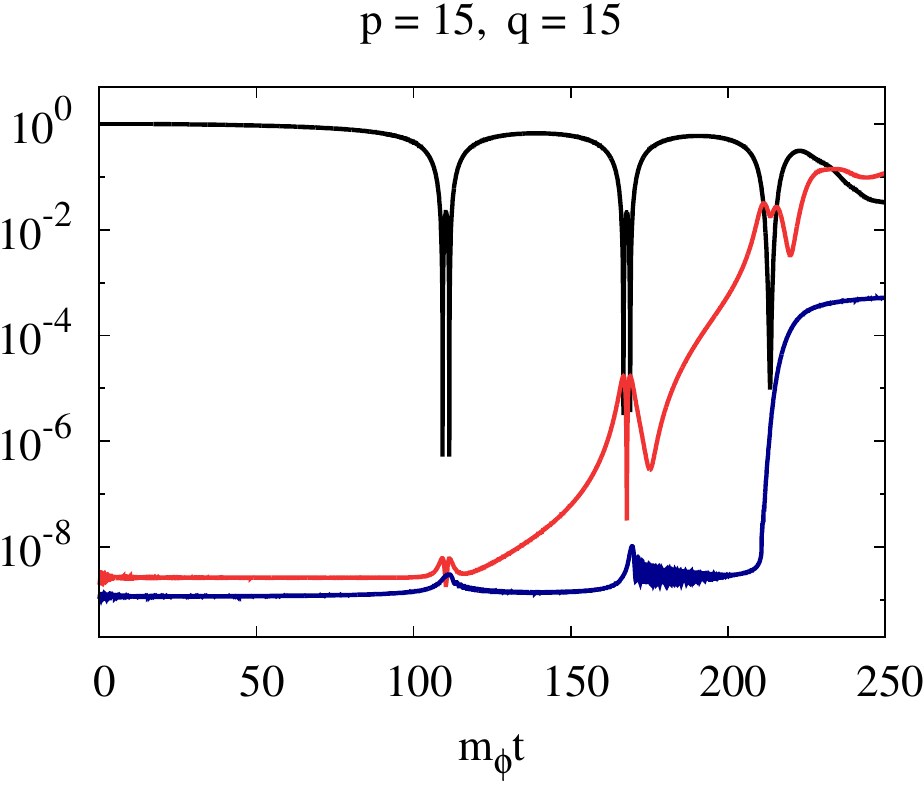}
	  \hspace{15mm}
	  \includegraphics[width=0.4\textwidth]{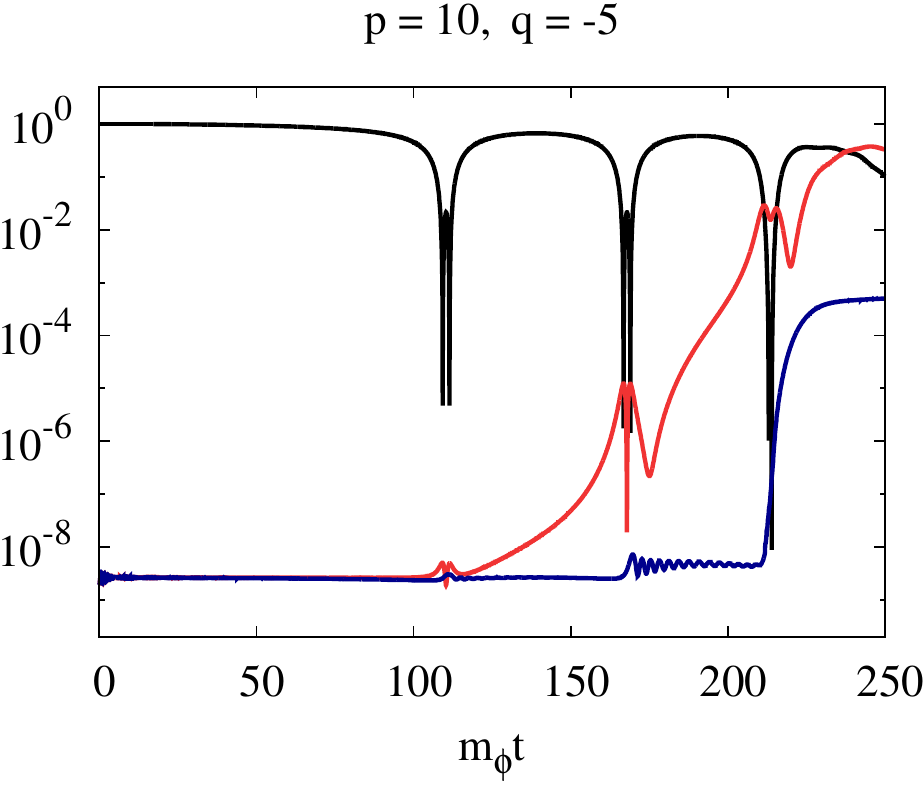}
	  \vspace{5mm}
	  
	  \includegraphics[width=0.4\textwidth]{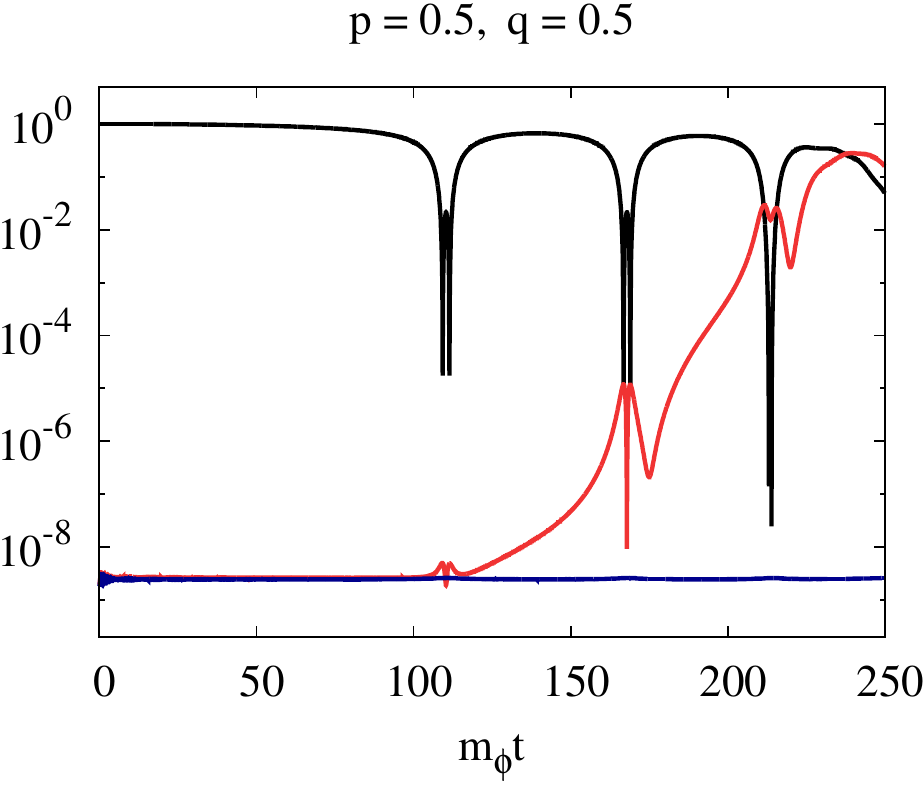}
	  \hspace{15mm}
	  \includegraphics[width=0.4\textwidth]{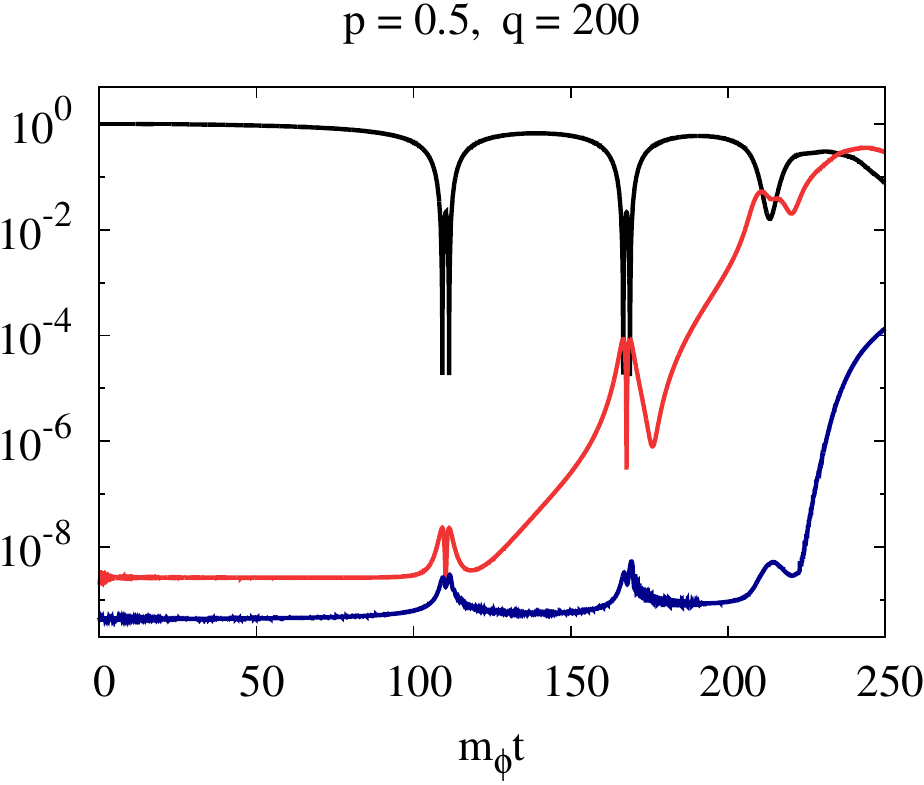}
\caption{\small
	The time evolution of the inflaton and Higgs for $v_\phi = 10^{-3}\Mpl$ up to $m_\phi t = 250$.
	The black line is the inflaton condensation $\langle \varphi \rangle^2$,
	\textcolor{light-red}{the red line} is the inflaton two point function 
	$\langle \varphi^2 \rangle - \langle \varphi \rangle^2$ and
	\textcolor{dark-blue}{the blue line} is the Higgs two point function $\langle h^2 \rangle$,
	where the angle brackets denote the spatial average.
	They are normalized by the initial inflaton amplitude $\varphi_\text{ini}$.
	The EW vacuum is stable for $(p, q) = (0.5, 0.5)$, while
	it is destabilized during the preheating for the other cases.
	The lower right panel corresponds to the case with an accidental cancellation
	between $\sigma_{\phi h}$ and $\lambda_{\phi h}$.
}
\label{fig:v10-3}
\end{figure*}

In this section we perform classical lattice simulations
to study the EW vacuum stability during the preheating epoch.
For concreteness, we take $n = 6$ in the inflaton potential~(\ref{Vinf}).
The CMB normalization~(\ref{Planck}) implies
\begin{align}
	\left(\frac{\Lambda}{\Mpl}\right)^4
	\simeq 
	7\times 10^{-14} \left(\frac{v_\phi}{\Mpl} \right)^3 \left(\frac{N}{60}\right)^{-5/2}.
\end{align}
For example, for $v_\phi/M_\text{Pl}=10^{-3}$, we have $\Lambda/M_\text{Pl} \simeq 3\times 10^{-6}$,
$H_\mathrm{inf} \simeq 10^7\,\mathrm{GeV}$ and $m_\phi \simeq 2\times 10^{11}\,\mathrm{GeV}$.
Thus the parameters satisfy $H_\mathrm{inf} \ll h_\mathrm{max} < m_\phi$,
and hence this model is indeed a good example of our general argument in the previous sections.
The condition (\ref{CW}) is given in terms of $p$ and $q$ as
\begin{align}
	\left\vert p\right\vert \lesssim \frac{v_\phi}{m_\phi}\left(\frac{v_\phi}{\Mpl}\right)^{5/4},
	~~~~~~~
	\left\vert q\right\vert \lesssim \frac{v_\phi}{m_\phi}\left(\frac{v_\phi}{\Mpl}\right)^{5/4}.
	\label{eq:CW_pq}
\end{align}
In the present case of $n=6$, the right-hand sides of these inequalities 
are larger than unity for $v_\phi/M_\text{Pl} \gtrsim 10^{-4}$.

We numerically solved the classical equations of motion derived from the Lagrangian~\eqref{eq:Lag}
as well as the Friedmann equations.
We start to solve the equations when the slow-roll parameter $\epsilon$ becomes unity.
It corresponds to $\cphi_\mathrm{ini} \simeq 0.74 v_\phi$ for $v_\phi = 10^{-2}\Mpl$, and
$\cphi_\mathrm{ini} \simeq 0.84 v_\phi$ for $v_\phi = 10^{-3}\Mpl$.
We took the initial velocity of the inflaton as zero.
We also introduced initial Gaussian fluctuations
that mimic the quantum fluctuations for the inflaton and the Higgs.
We have assumed that they are in the vacuum state initially.
This is justified for $v_\phi/\Mpl \gtrsim 10^{-6}$--$10^{-5}$
since we can safely neglect inflaton particle production at the first stage in this case
as discussed in Sec.~\ref{sec:inf}.
We have also added $h^6$ term in the Higgs potential just for numerical convergence.
We have checked that it does not modify the dynamics before the EW vacuum decays.
The parameters of our lattice simulations
are summarized in Tab.~\ref{tab:lattice}.
For more details on the classical lattice simulation, 
see for instance Refs.~\cite{Ema:2016kpf,Felder:2000hq,Frolov:2008hy} and references therein.

\begin{table}[t]
	\begin{tabular}{ccccc} \hline
	$d$ & $N_g$ & $L$ & $\mathrm{d}t$ & $\mathrm{d}k$ \\ \hline\hline
	~2+1 ~~&~~ 2048 ~~&~~ $1500 m_\phi^{-1}$ ~~&~~ $5\times10^{-3}m_\phi^{-1}$ 
	~~&~~ $4.2\times 10^{-3}m_\phi$ \\ \hline
	\end{tabular}
	\caption{\small The parameters of our lattice simulation, where 
	$d$ is the spacetime dimension, $N_g$ is the number of grid in each spatial dimension,
	$L$ is the size of the lattice, $\mathrm{d}t$ is the size of the each time step,
	$\mathrm{d}k \equiv 2\pi/L$ is the resolution of the momentum.}
	\label{tab:lattice}
\end{table}

Since we have two different momentum scales (Eq.~\eqref{eq:p_inf} and $m_\phi$),
we must take the number of grids $N_g$ to be large. 
This is why we took the spatial dimension to be two instead of three
(see Tab.~\ref{tab:lattice}).
As far as the linear regime is concerned, the results are not expected to change drastically
for different numbers of spatial dimensions.

We show our numerical results for $v_\phi = 10^{-2}\Mpl$ and $v_\phi = 10^{-3}\Mpl$ 
in Figs.~\ref{fig:v10-2} and~\ref{fig:v10-3} respectively.
We have followed the dynamics until $m_\phi t = 150$ and $250$ 
for $v_\phi = 10^{-2}\Mpl$ and $10^{-3}\Mpl$, respectively, since the inflaton condensation
is broken slightly before these times.
The black line is the inflaton condensation $\langle \varphi \rangle^2$,
\textcolor{light-red}{the red line} is the inflaton dispersion
$\langle \varphi^2 \rangle - \langle \varphi \rangle^2$, and
\textcolor{dark-blue}{the blue line} is the Higgs dispersion $\langle h^2 \rangle$,
where the angle brackets denote the spatial average. 
They are normalized by the initial amplitude of the inflaton condensation $\cphi_\mathrm{ini}^2$.
The resonance parameters $p$ and $q$ are written at the tops of these figures.

Let us start with the upper panels 
in Figs.~\ref{fig:v10-2} and~\ref{fig:v10-3}.
There, the resonance parameter $p$ satisfies 
$p \gtrsim \mathcal{O}(1)$, and
both $q > 0$ and $q < 0$ cases are considered.
As we can see from the figures, the EW vacuum is 
actually destabilized during the preheating for these cases.
On the other hand, we have taken the resonance parameters as 
$p = q \lesssim \mathcal{O}(1)$
in the lower left panels in Figs.~\ref{fig:v10-2} and~\ref{fig:v10-3}.
In these cases, the EW vacuum survives the preheating.
Thus the numerical results are consistent with our expectation in Sec.~\ref{sec:Higgs}. That is,
\begin{align}
	\left\lvert p \right\rvert = \frac{2\sigma_{\phi h}\cphi_\mathrm{ini}}{m_\phi^2} \lesssim \mathcal{O}(1),
	\label{eq:const_p}
\end{align}
is required for the stability of the EW vacuum during the preheating.
We have checked that this criterion is indeed satisfied 
for several other values of $p$ and $q$.
In particular, we have also calculated the case $p < 0$.
In this case, the Higgs becomes tachyonic in the region $\cphi > 0$,
where it takes more time for the inflaton to oscillate.
Hence the Higgs is more likely to be enhanced
and the EW vacuum decays faster compared to the case $p > 0$.\footnote{
	Note that the trilinear coupling eventually dominates over the quartic coupling
	as the inflaton approaches to the minimum of its potential.
} 
In any case, the EW vacuum is stable 
during the preheating as long as Eq.~\eqref{eq:const_p} holds
and $\vert q \vert \sim \vert p \vert$.
The bound \eqref{eq:const_p} does not strongly depend on $v_\phi$ 
since it is expressed solely by the resonance parameters.
It is consistent with the numerical results
with two different values of $v_\phi$.

Eq.~\eqref{eq:const_p} is our main result in this paper, 
and it also implies $\vert q \vert \lesssim \mathcal{O}(1)$ if there is no tuning of the parameters.
Still, we have also considered the case $\vert p\vert \ll q$ for the completeness of our study.
Note again that an accidental cancellation between $\sigma_{\phi h}$ and $\lambda_{\phi h}$ is 
necessary to achieve $q \gg \mathcal{O}(1)$ while satisfying Eq.~\eqref{eq:const_p}
(see the footnote~\ref{fn:q}).
In this case, the situation is more complicated. When the parametric resonance is dominant,
the condition for the EW vacuum destabilization in the linear regime 
is estimated as~\cite{Ema:2016kpf,Enqvist:2016mqj}\footnote{
	Apparently, the condition, $|\lambda_h \langle h^2 \rangle| \lesssim k_h^2$, 
	does not guarantee the stability for the homogeneous mode of the Higgs,
	but actually it does. We briefly explain the reason below. 
	See Ref.~\cite{Ema:2016kpf} for the original argument.
	As can be seen from Eq.~\eqref{eq:WH}, the Higgs acquires a positive mass term 
	from the Higgs-inflaton coupling. The Higgs escapes from its origin 
	only when the tachyonic mass, $| \delta m^2_\text{tac;h} |$, 
	overcomes the Higgs inflaton coupling. Expanding the effective Higgs mass around $\phi = v_\phi$, 
	one can estimate the time interval, $\delta t$, 
	during which $| \delta m^2_\text{tac;h} | \gtrsim m_h^2 (\phi)$ 
	as $| \delta m^2_\text{tac;h} | \sim q m_\phi^4  \delta t^2$.
	If the tachyonic mass term significantly drives the Higgs field during this time interval,
	or $|\delta m^2_\text{tac;h}| \delta t \gtrsim 1$,
	the vacuum decay takes place.
	This requirement coincides with Eq.~\eqref{eq:cond_q}.
}
\begin{align}
	\left\vert \lambda_h \right\vert \left\langle h^2 \right\rangle \gtrsim k_{h}^2,
	\label{eq:cond_q}
\end{align}
where $k_{h} \equiv m_\phi q^{1/4}$ is the typical momentum of the produced Higgs particles.
The dispersion grows like $\langle h^2 \rangle \sim k_h^2 e^{\mu_g m_\phi t}$ and the growth factor $\mu_g$ does not much
depend on $q$ for the parametric resonance~\cite{Kofman:1997yn}.
Hence the value of $q$ is not so important in this condition.
As a result, it is likely that the EW vacuum does not decay during the linear regime
even if we take $q$ to be larger,
since we have restricted the number of times of the inflaton oscillations in our analysis (only several times) 
to avoid complications associated with the nonlinear behavior of the inflaton.
However, as the inflaton fluctuations grow and become nonlinear, 
they can also produce the Higgs particles through the scatterings.
It corresponds to the beginning of the thermalization, 
which is studied in detail in, \textit{e.g.}, Ref.~\cite{Podolsky:2005bw}.
In this regime,
the variance of the fields interacting with each other
tends to converge to a similar value though the scattering.
Therefore, as $q$ (or $\lambda_{\phi h}$) becomes larger, 
the variance of the Higgs particles $\langle h^2 \rangle$ 
approaches to that of the inflaton $\langle \cphi^2\rangle$ faster.
In the present case, it might destabilize the EW vacuum
since $\vert \lambda_h \rvert \gg \lambda_{\phi h}$.
Actually, in the lower right panels in Figs.~\ref{fig:v10-2} and~\ref{fig:v10-3},
the EW vacuum is destabilized at almost the same time as the system becomes nonlinear
for $q \gtrsim \mathcal{O}(10)$.
Thus, it might be expected that
\begin{align}
	q = \frac{\lambda_{\phi h}\cphi_\mathrm{ini}^2}{m_\phi^2} \lesssim \mathcal{O}(10)
	~~~~
	\mathrm{if}
	~~~
	\left\vert p \right\vert \lesssim \mathcal{O}(1),
	\label{eq:const_q}
\end{align}
is at least required for the stability of the EW vacuum during and also after the preheating.

If we follow the thermalization process for a longer time,
the constraints may become tighter than Eqs.~\eqref{eq:const_p} and~\eqref{eq:const_q}.
In this sense, 
Eqs.~\eqref{eq:const_p} and~\eqref{eq:const_q} are just necessary conditions,
and we must also follow the dynamics after the preheating 
to determine an ultimate fate of the EW vacuum.
However, to address this issue, 
we should take into account the couplings between 
the Higgs and the SM particles, which might stabilize the EW vacuum.
We leave such a study for future work.

\section{Summary and discussions}
\label{sec:sum}

In this paper, we have studied the implications of the EW vacuum metastability
during the preheating epoch with low-scale inflation models, taking a hilltop inflation model as an example.
We have shown that, although the EW vacuum is naturally stable during inflation for low-scale inflation models,
it may decay into the deeper minimum during the preheating epoch due to the resonant Higgs production.

One of the particular features of the hilltop inflation model is that there is a tachyonic preheating
in the inflaton sector itself, which is so strong that the inflaton fluctuation becomes nonlinear within several inflaton oscillations.
To avoid complications arising from the nonlinearity of the inflaton,
we derive necessary conditions of the resonance parameters
as $|p| \lesssim \mathcal O(1)$ and $q \lesssim \mathcal{O}(10)$
by requiring that the vacuum remains stable until the inflaton becomes nonlinear (see Eq.~\eqref{eq:res_param} for the definitions of $p$ and $q$).
However, we also find that even after the inflaton field becomes completely inhomogeneous,
thermalization processes between the inflaton and Higgs tend to enhance the Higgs fluctuation,
which might cause the EW vacuum decay.
In addition to that, the production of other SM particles may also become relevant
for such a long time scale,
whose effects are unclear.
We did not give concrete bound taking into account such effects
due to the complexity of the system and limitation of the numerical simulation.
In this sense, the bounds we derived should be regarded as just a necessary condition.

Still it might be possible to estimate 
\textit{sufficient} conditions on the Higgs-inflaton couplings to avoid the EW vacuum decay.
If the couplings are small enough ($|p|, |q| \ll 1$), the band width of the Higgs resonance becomes narrow~\cite{Dolgov:1989us,Traschen:1990sw,Shtanov:1994ce}
and the Hubble expansion can kill the resonant Higgs production.
The condition that the narrow resonance does not happen is written as~\cite{Kofman:1997yn},
\begin{align}
	p^2,\,q^2 \lesssim \frac{H_{\rm inf}}{m_\phi} \sim \frac{v_\phi}{\Mpl}.  \label{pq_narrow}
\end{align}
If it is satisfied, the only way to produce Higgs bosons is the ordinary perturbative decay/annihilation of the inflaton
(without Bose enhancement).
The perturbative decay/annihilation rate may be estimated as
\begin{align}
	\Gamma(\varphi\to hh) \simeq \frac{\sigma_{\phi h}^2}{32 \pi m_\phi},~~~~~~
	\Gamma(\varphi\varphi\to hh) \sim \frac{\lambda_{\phi h}^2 \left<\varphi^2\right>}{32 \pi m_\phi}.
\end{align}
One may estimate the conservative bound which is free from the uncertainty of thermalization,
by requiring that the Higgs dispersion from the perturbative decay/annihilation 
never exceeds the instability scale, $\langle h^2 \rangle < h_\text{inst}^2$:
\begin{align}
	p^2, q^2 \lesssim \mathcal O (10^2) \frac{v_\phi}{\Mpl} \frac{h_\text{inst}^2}{m_\phi^2}. \label{eq:conservative}
\end{align}
While the bounds (\ref{eq:conservative}) might be too conservative,
it should be noted that we need to take account of the whole thermalization process including gauge bosons and quarks
in order to derive more precise bounds.

There are few remarks. 
First, we would like to 
comment on possible interactions between the Higgs and the inflaton
that are not taken into account in the main text.
Although it is higher dimensional, the following term can be large
for it respects the shift symmetry of inflaton:
\begin{align}
	\delta \mathcal L_\text{kin} = c_\text{kin}\frac{h^2}{M_\text{Pl}^2} \left( \der \phi \right)^2.
\end{align}
It induces an oscillating Higgs effective mass during the preheating,
and hence excites the Higgs fluctuations.
If we use the crude approximation that the inflaton potential is quadratic at around the minimum,
this coupling contributes to $A$ and $q$ in addition to $\lambda_{\phi h}$, 
making them independent even for the mode $k = 0$.
By requiring again Eq.~\eqref{eq:const_q}, we roughly estimate the constraint as
\begin{align}
	\left\vert c_\mathrm{kin}\right\rvert \lesssim \mathcal{O}(10)\times\frac{\Mpl^2}{v_\phi^2}.
\end{align}
In Ref.~\cite{Ema:2017loe}, it is found that the resonance can be suppressed
by making the ratio $A/q$ to be larger. However, in the present case, 
the inflaton potential is actually far from quadratic just after inflation,
and hence it might be difficult to cancel the oscillating part between $\dot{\cphi}^2$ and $\cphi^2$.
A similar discussion can be applied for the Higgs-gravity non-minimal coupling $\xi_h h^2 R$.

The next one is the possibility that the Higgs mass at $\cphi = 0$ in the early universe 
is different from that in the present universe.
It is possible if, for instance, the Higgs couples to a scalar field $\chi$ other than the inflaton
which has a finite VEV in the early universe.
The cancellation~\eqref{eq:cancel} does not hold in this case, 
and the resonance due to the inflaton oscillation can be suppressed 
if the Higgs mass at $\cphi = 0$ is larger than of order $m_\phi$.
However, $\chi$ must relax to its potential minimum at some epoch
so that the Higgs mass is of the order of the EW scale in the present universe.
We may need to discuss the resonant Higgs production during such a relaxation of $\chi$ instead,
if the mass of $\chi$ is larger than the instability scale of the EW vacuum.

Third, we comment on other low-scale inflation models.
While there are various class of low-scale inflation models, we expect that the bounds we found ($|p| \lesssim \mathcal O(1)$
and $|q| \lesssim \mathcal O(10)$) do not change much.
This is because our bounds only depend on the form of the
Higgs-inflaton potential around the minimum~(\ref{eq:pot_min}).
Thus they may be applied to other low-scale models \textit{e.g.}, hybrid inflation~\cite{Linde:1991km} and attractor inflation~\cite{Kallosh:2013yoa}, although more detailed study is needed to rigorously confirm it.

Finally, we again stress that, 
it is still far from clear in what condition
the EW vacuum is stable from the end of the preheating to the end of the thermalization process.
On the one hand, the EW vacuum stability during inflation and preheating is studied in detail 
in this paper as well as the previous literature~\cite{Espinosa:2007qp,Lebedev:2012sy,Kobakhidze:2013tn,
Fairbairn:2014zia,Enqvist:2014bua,Hook:2014uia,
Herranen:2014cua,Kamada:2014ufa,Shkerin:2015exa,Kearney:2015vba,Espinosa:2015qea,
East:2016anr,Joti:2017fwe,
Herranen:2015ima,Ema:2016kpf,Kohri:2016wof,Enqvist:2016mqj,Postma:2017hbk,Ema:2017loe}. 
On the other hand, it is known that the lifetime of the EW vacuum is long enough 
once the system is completely thermalized~\cite{Espinosa:2007qp,Rose:2015lna}.
However, we are still lacking studies on the EW vacuum (in)stability 
from the end of the preheating to the end of the thermalization.
Just after the preheating,
the momentum distribution of the Higgs (as well as the other SM particles)
is far from the thermal equilibrium,
and it evolves with time due to the scatterings
while approaching to the thermal equilibrium.
It is possible that the EW vacuum decay is activated during this thermalization process
depending on the shape of the momentum distribution.
For instance, if the Higgs modes become larger than other SM particles at some time, 
the vacuum decay can be enhanced; 
the resonant particle production studied in this paper may be viewed 
as an extreme example of this situation.
Thus, it is expected that the fate of our vacuum strongly depends on the detailed thermalization process.
Moreover, the process of thermalization depends also on the reheating temperature of the universe,
namely the coupling between the inflaton and SM particles other than the Higgs.
This issue is worth investigating in detail since we cannot avoid discussions on this point 
to determine an ultimate fate of the EW vacuum.
Hopefully we will come back to this issue in future publication.

\section*{Acknowledgments}
\small
This work was supported in part by the JSPS Research Fellowships for Young Scientists (YE and KM)
and the Program for Leading Graduate Schools, MEXT, Japan (YE).
This work was also supported by the Grant-in-Aid for Scientific Research on Scientific
Research A (No.26247042 [KN]), Young Scientists B
(No.26800121 [KN]) and Innovative Areas (No.26104009 [KN], No.15H05888 [KN]).

\appendix

\section{Tachyonic preheating after hilltop inflation} \label{sec:app}

In this Appendix, we summarize the properties of tachyonic preheating during the inflaton oscillation after hilltop inflation.
The most discussion below follows Ref.~\cite{Brax:2010ai}.

Let us denote by $\phi_j$ the lower endpoint field value of the inflaton after $j$-th inflaton oscillation
and $t_j$ the time at which $\phi=\phi_j$.
The endpoint is evaluated from the energy conservation:
\begin{align}
	V(\phi_j) - V(\phi_{j+1}) =\int_{t_j}^{t_{j+1}}dt\,3H\dot\phi^2 = \int_{\phi_j}^{\phi_{j+1}}d\phi\,3H\dot\phi.
\end{align}
The integral is dominated around the potential minimum where $\phi \sim v$ and $|\dot\phi| \sim \Lambda^2$.
Thus we obtain
\begin{align}
	\frac{\phi_j}{v_\phi} \sim \left(\frac{j\sqrt 3}{2} \frac{v_\phi}{\Mpl}\right)^{1/n}.
\end{align}
Note that $\phi_{j=1} > \phi_{\rm end}^{(\epsilon)}$, where $\phi_{\rm end}^{(\epsilon)}/v_\phi \sim (v_\phi/\Mpl)^{1/(n-1)}$ denotes the field value at
$\epsilon=1$.
The time period of $j$-th oscillation is given by
\begin{align}
	t_{j+1} - t_j \sim \frac{1}{m_\phi}\left(\frac{v_\phi}{\phi_j}\right)^{(n-2)/2},
\end{align}
hence it is much longer than the inverse of the mass scale around the potential minimum,
as clearly seen in Figs.~\ref{fig:v10-2}-\ref{fig:v10-3}.

We consider the growth of the inflaton fluctuation $\delta\phi_k$ with a wavenumber $k$ 
in the linear approximation
during $j+1$ oscillation: $t_j \leq t \leq t_{j+1}$.
Below, we neglect the Hubble expansion since all the time scales are much shorter than the Hubble time scale
and take the scale factor $a=1$ for notational simplicity.
We further divide the one oscillation into three phases: (a) $t_j < t < t_m^-$, (b) $t_m^-<t<t_m^+$, (c) $t_m^+ < t < t_{j+1}$,
where $t_m^\pm$ denotes the time when $\phi$ passes through $\phi_m$,
the field value at which $V''$ takes negative maximum value:
\begin{align}
	\frac{\phi_m}{v_\phi} = \left( \frac{n-2}{2(2n-1)} \right)^{1/n}.
\end{align}
First, in the stage (a), modes with $k \lesssim m_\phi$ experience tachyonic instability within the field range $\phi_{\rm tac} \leq \phi < \phi_m$, where $\phi_{\rm tac}$ at which the mode begins to be tachyonic:
\begin{align}
	\frac{\phi_{\rm tac}}{v_\phi} \simeq\frac{\phi_j}{v_\phi}\times {\rm max}\left[1,~\left(\frac{k}{k_*} \right)^{2/(n-2)}\right],
\end{align}
with $k_* \sim m_\phi(jv_\phi/M_P)^{(n-2)/(2n)}$ corresponding to the tachyonic mass scale around $\phi=\phi_j$.
Then the inflaton fluctuation $\delta\phi_k$ is enhanced by an exponential factor $e^{X_k}$ with
\begin{align}
	X_k = \int_{\phi_{\rm tac}}^{\phi_m}\frac{\sqrt{|V'' + k^2|}}{\dot\phi}d\phi \sim \sqrt{\frac{n(n-1)}{2}}\log\left( \frac{\phi_m}{\phi_{\rm tac}} \right).   \label{Xk}
\end{align}
However, it should be noticed that the same mode also experiences exponential decay in the third stage (c).
It is easy to imagine that in the limit of $k\to 0$ this exponential decay during the stage (c) exactly cancels the exponential growth during the stage (a), because it is just the same as the dynamics of the homogeneous mode.
For finite $k$, however, there is a phase shift during the stage (b), which causes a mismatch between the growing solution in the stage (a)
and the decaying solution in the stage (c).
Schematically, the phase of $\delta\phi_k$ is rotated during the stage (b) as
\begin{align}
	e^{i\sqrt{m_\phi^2+k^2} t} \sim e^{im_\phi t}\left(1+i\frac{k^2}{m_\phi^2}\right),
\end{align}
where we used $k\ll m_\phi$ and $t\sim t_m^+-t_m^- \sim m_\phi^{-1}$.
Therefore, a small fraction of $k^2/m_\phi^2$ at the end of stage (b) connects to the growing mode in the stage (c).
The net enhancement factor in one oscillation is then estimated as
\begin{align}
	F_k\equiv\left|\frac{\delta\phi_k(t_{j+1})}{\delta\phi_k(t_j)}\right|\sim \left|1+ i\frac{k^2}{m_\phi^2}e^{2X_k}\right|.
\end{align}
 Using (\ref{Xk}), it is found that $F_k$ is peaked around $k\simeq k_*$, where we have
\begin{align}
	F_{k_*} \sim \left( \frac{m_\phi^2}{k_*^2} \right)^{x_n-1},~~~~~~x_n\equiv \frac{\sqrt{2n(n-1)}}{n-2}.
\end{align}
Note that it is much larger than unity, hence the inflaton fluctuation is enhanced by orders of magnitude within one oscillation
for $v_\phi\ll \Mpl$.
This is much different from the ordinary preheating with the parametric resonance.

The variance of the field fluctuation after the $j$-th inflaton oscillation 
is now dominated by the modes $k\sim k_*$ 
and estimated as
\begin{align}
	\left<\delta\phi^2\right> \sim k_*^2(F_{k_*})^{2j} \sim m_\phi^2 \left( \frac{M_P}{v_\phi} \right)^{\frac{n-2}{n}\left[2j(x_n-1)-1 \right]}.
\end{align}
It is true if $x_n > 3/2$ which is valid for $n<27$.
Thus it will take only a few or several inflaton oscillations for the fluctuation to become nonlinear for low-scale inflation $v_\phi\ll \Mpl$.

\bibliographystyle{apsrev4-1}
\bibliography{ref}

\end{document}